\documentclass[prd,twocolumn,floatfix,amsmath,nofootinbib,amssymb,floatfix]{revtex4}
\usepackage{graphicx,color,dcolumn,booktabs,bm,multirow}
\usepackage{longtable,lscape}
\usepackage{txfonts}
\usepackage{overpic}
\usepackage{amssymb}
\usepackage{array}
\usepackage{indentfirst}
\usepackage{feynmf}   
\usepackage{slashed}  
\usepackage{cases}
\usepackage{color}
\usepackage{multirow}
\usepackage{epstopdf}
\usepackage[utf8]{inputenc}
\usepackage{graphicx,color,dcolumn,booktabs,bm}
\usepackage[colorlinks,
citecolor=blue,
anchorcolor=red,
menucolor=red,
linkcolor=red,
filecolor=red,
runcolor=red,
urlcolor=blue,
frenchlinks=red]{hyperref}
\usepackage{float}
\usepackage{mathrsfs}

\begin{document}

\title{$P_{c}(4312)$, $P_{c}(4440)$, and $P_{c}(4457)$ productions in $e^{+} e^{-}$ collisions}
	\author{Quan-Yun Guo$^{1}$}
	\author{Dian-Yong Chen$^{1,2}$\footnote{Corresponding author}}\email{chendy@seu.edu.cn}
	\affiliation{
$^1$ School of Physics, Southeast University, Nanjing 210094, China\\
$^2$Lanzhou Center for Theoretical Physics, Lanzhou University, Lanzhou 730000,China\\
	}
\date{\today}

\begin{abstract}
In the present work, we propose to investigate the productions of $P_{c}(4312)$, $P_{c}(4440)$, and $P_{c}(4457)$ in the $e^{+} e^{-} \rightarrow p \bar{p} J/\psi$ process by utilizing an effective Lagrangian approach, where $P_{c}(4312)$ is considered as $\Sigma_{c} \bar{D}$ molecular state with $J^{P}=1/2^{-}$, while $P_{c}(4440)$ and $P_{c}(4457)$ are considered as $\Sigma_{c} \bar{D}^{\ast}$ molecular states with $J^{P}=1/2^{-}$ and $J^{P}=3/2^{-}$, respectively. For the $e^{+} e^{-} \rightarrow \bar{p} P_{c}(4312)$, $e^{+} e^{-} \rightarrow \bar{p} P_{c}(4440)$, and $e^{+} e^{-} \rightarrow \bar{p} P_{c}(4457)$ processes, the cross sections are evaluated to be ($27.8^{+42.2}_{-17.9}$) fb, ($38.7^{+58.6}_{-25.1}$) fb, and ($1.44^{+2.18}_{-0.93}$) fb at $\sqrt{s}=6$ GeV, respectively, where the central values are estimated with $\Lambda_{r}=2.5$ GeV, and the uncertainties are resulted from the variations of $\Lambda_{r}$ from 2.0 to 3.0 GeV. Considering that the $P_{c}$ states can decay into $J/\psi p$, we estimate the cross sections for $e^{+} e^{-} \rightarrow p \bar{p} J/\psi$ and the differential cross sections depending on the $J/\psi p$ invariant mass. At $\sqrt{s}=6$ GeV, the cross sections for $e^{+} e^{-} \rightarrow p \bar{p} J/\psi$ are estimated to be ($14.7^{+22.2}_{-9.50}$) fb. Moreover, our estimations indicate that the cross section resulted from the $P_{c}(4440)$ and $\bar{P}_{c}(4440)$ intermediate states is dominant. In the $J/\psi p$ invariant mass spectrum, our estimations reveal that the peak structure between 4.42 and 4.46 GeV primarily originates from $P_{c}(4440)$.
\end{abstract}

\maketitle

\section{Introduction}
\label{sec:Introduction}
Since the observation of $X(3872)$ in 2003~\cite{Belle:2003nnu}, the studies of multiquark candidates have advanced rapidly (See Refs.~\cite{Belle:2007hrb, Belle:2013yex, BESIII:2013ouc, BaBar:2006itc, Belle:2006xni, Belle:2022hnm, BaBar:2003oey, CLEO:2003ggt, Guo:2017jvc, Ali:2017jda, Olsen:2017bmm, Liu:2019zoy, Bai:2026atm} for representative examples). In 2015, the potential pentaquark states, $P_{c}(4380)$ and $P_{c}(4450)$, were observed in the $J/\psi p$ invariant mass distributions of the $\Lambda^{0}_{b} \rightarrow J/\psi K^{-} p$ decay process by the LHCb Collaboration~\cite{LHCb:2015yax, LHCb:2016ztz, LHCb:2016lve}. In 2019, the LHCb Collaboration reanalyzed the $\Lambda^{0}_{b} \rightarrow J/\psi K^{-} p$ process with more data, and reported a new narrow state, $P_{c}(4312)$, in the $J/\psi p$ invariant mass distributions with a significance of 7.3$\sigma$~\cite{LHCb:2019kea}. Moreover, the previously reported $P_{c}(4450)$ structure was confirmed to consist of two narrow structures corresponding to $P_{c}(4440)$ and $P_{c}(4457)$. Subsequent observations of the strange partner of $P_c$ states, namely $P_{cs}(4459)$ and $P_{cs}(4380)$ respectively observed in the $J/\psi \Lambda$ invariant mass spectrum of $\Xi^{-}_{b} \rightarrow J/\psi \Lambda K^{-}$ decay~\cite{LHCb:2020jpq} and $B^{-} \rightarrow J/\psi \Lambda \bar{p}$ decay~\cite{LHCb:2022ogu}, make the hidden charm pentaquark family abundant.

At present, the Particle Data Group averages for the resonance parameters of $P_{c}(4312)$, $P_{c}(4440)$, and $P_{c}(4457)$ are~\cite{ParticleDataGroup:2024cfk},
\begin{eqnarray}
	P_{c}(4312) : &\mathrm{M}&=\Big(4311.9^{+7.0}_{-0.9}\Big) \; \mathrm{MeV}, \nonumber\\ &\Gamma&=\Big(10 \pm 5\Big) \;  \mathrm{MeV}, \nonumber\\ P_{c}(4440) : &\mathrm{M}&=\Big(4440^{+4}_{-5}\Big) \; \mathrm{MeV}, \nonumber\\ &\Gamma&=\Big(21^{+10}_{-11}\Big) \; \mathrm{MeV}, \nonumber\\ P_{c}(4457) : &\mathrm{M}&=\Big(4457.3^{+4.0}_{-1.8}\Big) \; \mathrm{MeV}, \nonumber\\ &\Gamma&=\Big(6.4^{+6.0}_{-2.8}\Big) \; \mathrm{MeV}. \label{eq.1}
\end{eqnarray}
respectively.

The observed mass of $P_{c}(4312)$ is close to the threshold of $\Sigma_{c}\bar{D}$, while the masses of $P_{c}(4440)$ and $P_{c}(4457)$ are well consistent with the fine structure of $\Sigma_{c}\bar{D}^{\ast}$ interaction with the total spin $1/2$ and $3/2$, which suggests that they are likely $\Sigma_{c}\bar{D}$ and $\Sigma_{c}\bar{D}^{\ast}$ pentaquark molecular states, respectively. In Ref.~\cite{Chen:2015loa,He:2015cea,Chen:2019asm}, the authors investigated the $\Sigma_{c}\bar{D}^{(\ast)}$ interaction using the one-boson-exchange model. Their results indicate that $P_{c}(4312)$, $P_{c}(4440)$, and $P_{c}(4457)$ could be regarded as loosely $\Sigma_{c}\bar{D}$ bound state with $J^{P}=1/2^{-}$, $\Sigma_{c}\bar{D}^{\ast}$ bound state with $J^{P}=1/2^{-}$, and $\Sigma_{c}\bar{D}^{\ast}$ bound state with $J^{P}=3/2^{-}$, respectively. The same conclusion could be obtained using the  quasipotential Bethe-Salpeter equation approach~\cite{He:2019ify}. The estimations in Ref.~\cite{Liu:2019tjn} indicated that these three $P_c$ states can be naturally accommodated within a contact-range effective field theory description that incorporates heavy-quark spin symmetry in the $\Sigma_{c}\bar{D}^{(\ast)}$ molecular frame. The QCD sum rule investigations in Ref.~\cite{Chen:2015moa} indicated that these $P_c$ states could be identified as exotic hidden-charm pentaquarks composed of an anticharmed meson and a charmed baryon. Besides the molecular interpretations, the properties of these $P_c$ states have also been investigated within the compact pentaquark framework~\cite{Lebed:2015tna, Li:2015gta, Wang:2015epa, Brambilla:2025xma, Brambilla:2026ujo}. In addition to the exotic interpretations, these structures have also been associated with threshold effects arising from the rich thresholds of a charmonium and a baryon~\cite{Guo:2015umn, Liu:2015fea, Meissner:2015mza}. In Ref.~\cite{Nakamura:2021qvy}, the author proposed that the peak structures $P_c(4312)$, $P_c(4380)$, and $P_c(4457)$ can be described by the double triangle cusps and their interference with the common mechanisms, while $P_c(4440)$ is interpreted as a resonance with a significantly smaller width than the LHCb measurement. In the same theoretical framework, the authors in Ref.~\cite{Nakamura:2021dix} further proposed that the peak structure $P_c(4337)$ can be described by a $\Sigma_{c} \bar{D}$ one-loop threshold cusp. Moreover, they concluded that the structures $P_c(4312)$ and $P_c(4337)$ originate from different interference patterns between the $\Sigma_{c} \bar{D}$ and $\Lambda_{c}\bar{D}^{\ast}$ threshold cusps.

In terms of decay properties, the authors in Ref.~\cite{Xiao:2019mvs} estimated the widths of $P_{c} \rightarrow J/\psi p$ to be of order several MeV in the molecular scenario, where $P_{c}(4312)$ was assigned as $\Sigma_{c} \bar{D}$ molecular state with $J^{P}=1/2^{-}$, while $P_{c}(4440)$ and $P_{c}(4457)$ were assigned as $\Sigma_{c} \bar{D}^{\ast}$ molecular states with $J^{P}=1/2^{-}$ and $J^{P}=3/2^{-}$, respectively. Assuming $P_{c}(4312)$ to be $\Sigma_{c}\bar{D}$ molecular state with $J^{P}=1/2^{-}$, the authors in Ref.~\cite{Xu:2019zme} estimated its partial widths by using the QCD sum rules. Their results showed that the widths of the $\eta_{c}p$ and $J/\psi p$ channels are  $(5.54^{+0.75}_{-0.50})$ MeV and $(1.67^{+0.92}_{-0.56}$) MeV, respectively. In Ref.~\cite{Yang:2024nss}, the authors estimated the decay widths of $P_{c}(4440)$ and $P_{c}(4457)$ to $\bar{D} \Sigma_{c}$ and $\bar{D} \Lambda_{c}$ under two different spin-parity assignments. From the perspective of decay properties, their estimations favor the assignment that the $J^P$ quantum numbers of $P_{c}(4440)$ and $P_{c}(4457)$ are $1/2^-$ and $3/2^-$, respectively.

Besides the mass spectrum and decay properties, the production mechanisms of $P_{c}$ states in various processes have also been investigated. The authors in Ref.~\cite{Wu:2019rog} investigated the productions of $P_{c}(4312)$, $P_{c}(4440)$, and $P_{c}(4457)$ in $\Lambda_{b}$ decays, where $P_{c}(4312)$ was interpreted as $\Sigma_{c} \bar{D}$ molecular state with $J^{P}=1/2^{-}$, while $P_{c}(4440)$ and $P_{c}(4457)$ were interpreted as $\Sigma_{c} \bar{D}^{\ast}$ molecular states with $J^{P}=1/2^{-}$ and $J^{P}=3/2^{-}$, respectively. Their estimations indicated that the branching fractions of $\Lambda_{b} \to P_{c}K$ are of order $10^{-6}$. In the same molecular frame, the authors in Ref.~\cite{Wang:2015jsa,Wang:2019krd} estimated the cross sections for the $\gamma p \rightarrow J/\psi p$ process by using an effective Lagrangian approach, where the branching fractions of $P_c \to J/\psi p$ were assumed to be $5\%$ and $10\%$. Following the above molecular framework, the authors in Ref.~\cite{Wang:2019dsi} studied the productions of $P_{c}(4312)$, $P_{c}(4440)$, and $P_{c}(4457)$ in the $\pi^{-}p \rightarrow J/\psi n$ process by including the $s$-, $u$-, and $t$-channels. Their results indicated that the average cross section from $P_{c}(4312)$ is about 1.2 nb/100 MeV. In addition, the authors in Ref.~\cite{Liu:2021ojf} investigated the productions of $P_{c}$ states in association with $Z_{c}(3900)/Z_{cs}(3985)$ via $\pi p$ and $Kp$ scattering. Their results indicated that the cross sections for the $\pi p \rightarrow Z_{c}(3900) P_{c}(4312)/P_{c}(4440)/P_{c}(4457)$ processes can reach the order of 10 nb. More recently, the Belle II Collaboration~\cite{Belle-II:2026heb} studied the $e^{+} e^{-} \rightarrow p \bar{p} J/\psi$ process for the first time, although no clear structure is observed in the $p \bar{p} J/\psi$ cross section, they reported upper
limits of the cross sections from the production threshold up to 7 GeV.

Considering that the high energy photon involved in the photon productions of $P_c$ states comes from the high energy electron, i.e., the full process should be $e^- p\to e^- P_c\to e^- J/\psi p$. Then, the process $e^+ e^- \to P_c \bar{p}$ could be constructed to investigate $P_c$ states by crossing symmetry. Specifically, the electron and positron first annihilate into a virtual photon, and then the photon couples to the $\bar{p} P_{c}$ final states through an electromagnetic interaction as shown in Fig.~\ref{Fig.1}. In terms of experimental conditions, the Super Tau-Charm facility (STCF)~\cite{Peng:2020orp, Achasov:2023gey, Ai:2025xop} is designed to operate with a center-of-mass energy range of 2 to 7 GeV and a peak luminosity of 5.0$\times$ $10^{34}$ cm$^{-2}$s$^{-1}$, which translates to a yield of events more than 50 times greater than that of BEPCII~\cite{Yu:2016cof}. Therefore, we propose to investigate the productions of $P_{c}$ states via $e^{+}e^{-}$ collisions in this work, where $P_{c}(4312)$ is considered as $\Sigma_{c} \bar{D}$ molecular state with $J^{P}=1/2^{-}$, while $P_{c}(4440)$ and $P_{c}(4457)$ are considered as $\Sigma_{c} \bar{D}^{\ast}$ molecular states with $J^{P}=1/2^{-}$ and $J^{P}=3/2^{-}$, respectively. Recently, the authors in Ref.~\cite{Zhang:2025pfz} estimated the cross section for $e^{+} e^{-} \rightarrow p \bar{p} J/\psi$ at $\sqrt{s}$ from 6 to 7 GeV. Their results indicated that the cross section for $e^{+}e^{-} \rightarrow \bar{p} P_{c}$ is $\lesssim \mathcal{O}(0.1~\text{pb})$, while that for $e^{+} e^{-} \rightarrow p \bar{p} J/\psi$ is $\mathcal{O}(4~\text{fb})$. In the present work, we first estimate the cross sections for the $e^{+}e^{-} \rightarrow \bar{p} P_{c}(4312)/P_{c}(4440)/P_{c}(4457)$ processes. Then, considering that the $P_{c}$ states can decay into $J/\psi p$, we estimate the cross sections for $e^{+} e^{-} \rightarrow p \bar{p} J/\psi$, including the contributions from $P_{c}(4312)$, $P_{c}(4440)$, $P_{c}(4457)$, and their antiparticles, respectively. 

This work is organized as follows. After introduction, we present our estimations of the cross sections for $e^{+}e^{-} \rightarrow \bar{p} P_{c}(4312)/P_{c}(4440)/P_{c}(4457)$ and $e^{+} e^{-} \rightarrow p \bar{p} J/\psi$. In Section \ref{sec:MA}, the numerical results and related discussions of the cross sections are presented. The last section is devoted to a short summary.

\section{$P_{c}(4312)$/$P_{c}(4440)$/$P_{c}(4457)$ productions in the $e^{+} e^{-}$ collisions}
\label{sec:MS}
\subsection{Cross Sections for $e^{+} e^{-} \rightarrow \bar{p} P_{c}(4312)$/$P_{c}(4440)$/$P_{c}(4457)$}
\begin{figure}[t]
	\begin{tabular}{ccc}
		\centering
		\includegraphics[width=75mm]{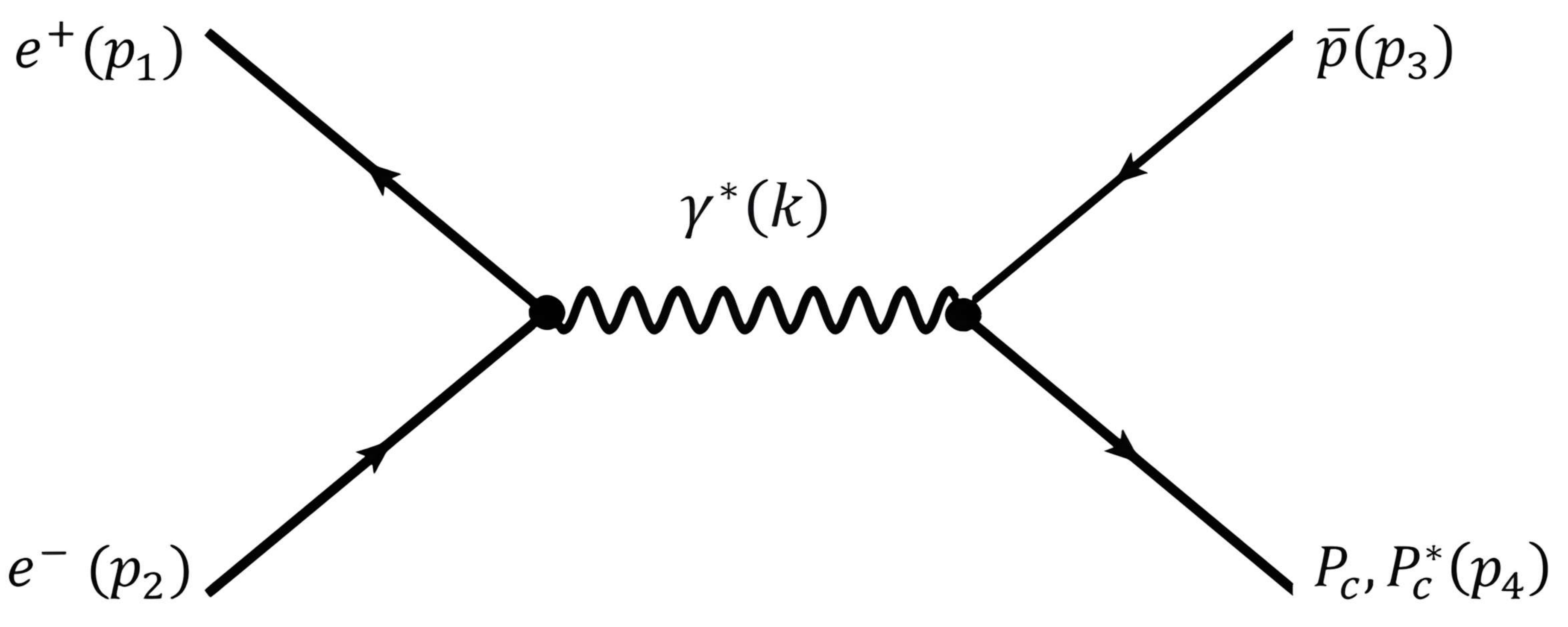}&\\
	\end{tabular}
	\caption{Diagram contributing to the $e^{+} e^{-} \rightarrow \bar{p} P^{(\ast)}_{c}$ process. Here $P_{c}$ and $P^{\ast}_{c}$ refer to $P_{c}(4312)$ and $P_{c}(4440)$/$P_{c}(4457)$, respectively.}\label{Fig.1}
\end{figure}

\begin{figure}[t]
	\begin{tabular}{ccc}
		\centering
		\includegraphics[width=80mm]{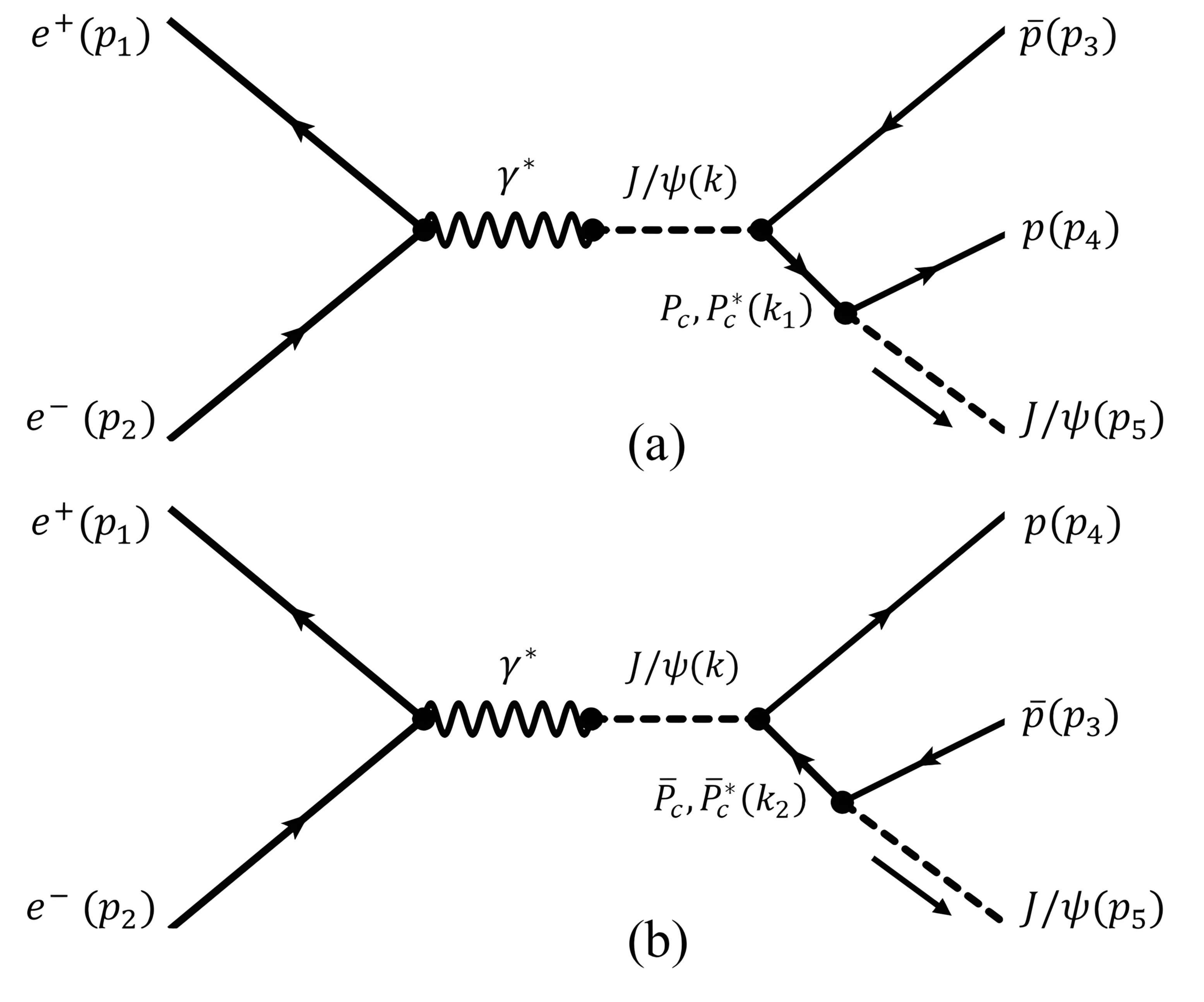}&\\
	\end{tabular}
	\caption{Diagrams contributing to the process of $e^{+} e^{-} \rightarrow p \bar{p} J/\psi$. Diagrams (a) and (b) correspond to the $P^{(\ast)}_{c}$ and $\bar{P}^{(\ast)}_{c}$ exchanges, respectively. Here $P_{c}$ and $P^{\ast}_{c}$ refer to $P_{c}(4312)$ and $P_{c}(4440)$/$P_{c}(4457)$, respectively, while $\bar{P}^{(\ast)}_{c}$ refer to the antiparticle of $P^{(\ast)}_{c}$.}\label{Fig.2}
\end{figure}
In Fig.~\ref{Fig.1}, we present the Feynman diagrams for $e^{+} e^{-} \rightarrow \bar{p} P_{c}(4312)/P_{c}(4440)/P_{c}(4457)$. In the present calculations, we employ the effective Lagrangian approach to depict the hadron interaction vertices. The effective Lagrangians for the $\gamma N P_{c}$ and $P_{c} \psi N$ vertices can be written as~\cite{Wang:2019krd,Wang:2019dsi,Liu:2021ojf, Huang:2016tcr},
\begin{eqnarray}
\mathcal{L}^{1/2-}_{\gamma N P_{c}}&=&\frac{eh}{2m_{N}} \bar{N} \sigma_{\mu \nu} \partial^{\nu} A^{\mu} P_{c} + \mathrm{H.c}.,\nonumber\\
\mathcal{L}^{3/2-}_{\gamma N P_{c}}&=&\frac{ieh_{1}}{2m_{N}} \bar{N} \gamma_{\nu} F^{\mu\nu} P_{c\mu} + \mathrm{H.c}.,\nonumber\\
\mathcal{L}^{1/2-}_{P_{c} \psi N}&=&g^{1/2-}_{P_{c} \psi N} \bar{N} \gamma_{5} \gamma_{\mu} P_{c} \psi^{\mu} + \mathrm{H.c}., \nonumber\\ 
\mathcal{L}^{3/2-}_{P_{c} \psi N}&=&g^{3/2-}_{P_{c} \psi N} \bar{N} \psi^{\mu} P_{c\mu} + \mathrm{H.c}., \label{eq.2}
\end{eqnarray}
where $F^{\mu \nu}=\partial^{\mu} A^{\nu} - \partial^{\nu} A^{\mu}$ and $\psi^{\mu \nu}=\partial^{\mu} \psi^{\nu} -\partial^{\nu} \psi^{\mu}$, respectively. The superscripts $1/2^{-}$ and $3/2^{-}$ correspond to the $J^{P}$ quantum numbers of the $P_{c}(4312)$/$P_{c}(4440)$ and $P_{c}(4457)$ final states, respectively. The electromagnetic coupling constants $eh$ and $eh_{1}$ can be obtained by the strong coupling constants $g^{1/2-}_{P_{c} \psi N}$ and $g^{3/2-}_{P_{c} \psi N}$ within the framework of vector meson dominance mechanism (VMD)~\cite{Bauer:1975bv, Bauer:1975bw, Bauer:1977iq}. Notably, the $P_{c}$ states have been reconstructed exclusively in the $J/\psi p$ final states, and their couplings to higher charmonia remain largely unconstrained by existing experimental data. Therefore, we focus on the $J/\psi$-mediated contribution in the present work. The effective Lagrangian for the $J/\psi$-photon coupling is
\begin{eqnarray}
\mathcal{L}_{J/ \psi \gamma}=-\frac{e m^{2}_{\psi}}{f_{\psi}} V_{\mu} A^{\mu}, \label{eq.4}
\end{eqnarray}
where $e$ is the electromagnetic coupling constant. $V_{\mu}$ and $A^{\mu}$ denote the $J/\psi$ meson and photon fields, respectively. Moreover, $f_{\psi}$ is the decay constant of $J/\psi$, its specific value will be discussed in Section \ref{sec:MA}. By comparing the radiative decay width estimated from directly $\gamma NP_c$ coupling and VMD framework, one has,
\begin{eqnarray}
eh&=&g^{1/2^{-}}_{P_{c} \psi N} \frac{e}{f_{\psi}} \frac{2 m_{N}}{(m^{2}_{P_{c}}-m^{2}_{N}) m_{\psi}} \nonumber\\&\times& \sqrt{m^{2}_{\psi}(m^{2}_{N}+4m_{N}m_{P_{c}}+m^{2}_{P_{c}})+(m^{2}_{P_{c}}-m^{2}_{N})^2},\nonumber\\
eh_{1}&=&g^{3/2^{-}}_{P_{c} \psi N} \frac{e}{f_{\psi}} \frac{2 m_{N}(m_{N}+m_{P_{c}})}{(m^{2}_{P_{c}}-m^{2}_{N})m_{\psi}}\nonumber\\&\times& \sqrt{\frac{6m^{2}_{\psi} m^{2}_{P_{c}}+m^{4}_{N}-2m^{2}_{N}m^{2}_{P_{c}} +m^{4}_{P_{c}}}{3 m^{2}_{P_{c}}+m^{2}_{N}}}.
\label{eq.3}
\end{eqnarray}

With the above effective Lagrangians, one can obtain the amplitudes corresponding to the $e^{+} e^{-} \rightarrow \bar{p} P_{c}(4312)$/$P_{c}(4440)$/$P_{c}(4457)$ processes, which are,
\begin{eqnarray}
\mathcal{M}^{1/2^{-}}&=&\Big[\bar{u}(p_{4}) \Big(\frac{eh}{2m_{N}} \sigma_{\mu \nu} (ik^{\nu})\Big) v(p_{3})\Big] \frac{-g^{\mu \alpha}}{k^{2}} \nonumber\\ &\times& \Big[\bar{v}(p_{1}) (ie \gamma^{\alpha}) u(p_{2}) \Big] \times F(k,m_{\gamma},\Lambda_{r}),\nonumber\\
\mathcal{M}^{3/2^{-}}&=&\Big[\bar{u}^{\mu}(p_{4}) \Big(\frac{ieh_{1}}{2m_{N}} \gamma^{\nu} (ik^{\mu}g_{\nu \rho}-ik^{\nu}g_{\mu \rho}) \Big) v(p_{3}) \Big] \nonumber\\ &\times& \frac{-g^{\rho \alpha}}{k^{2}} \Big[\bar{v}(p_{1}) (ie \gamma^{\alpha}) u(p_{2}) \Big] \times F(k,m_{\gamma},\Lambda_{r}). \label{eq.5}
\end{eqnarray}
where the superscripts $1/2^{-}$ and $3/2^{-}$ correspond to the $J^{P}$ quantum numbers of the $P_{c}$ in final states. In addition, the form factor $F (k_{i},m_{i}, \Lambda_{r})$ is introduced to depict the inner structure of the involved hadrons and the off-shell effects, and its specific form is,
\begin{eqnarray}
F(k_{i},m_{i},\Lambda_{r})
&=&\frac{\Lambda^{2}_{r}-m^{2}_{i}} {\Lambda^{2} _{r} - k^{2}_{i}},\label{Eq.6}
\end{eqnarray}
where $k_{i}$ and $m_{i}$ are the four momentum and the mass of the exchanged state, respectively. $\Lambda_{r}$ is a model parameter, and its specific value will be discussed in next section.

\subsection{Cross Sections for $e^{+} e^{-} \rightarrow p \bar{p} J/\psi$}
Since $P_{c}(4312)$, $P_{c}(4440)$, and $P_{c}(4457)$ can decay into $J/\psi p$ final states, we further investigate the $P_{c}$ states in the $e^{+} e^{-} \rightarrow p \bar{p} J/\psi$ process. The diagrams contributing to $e^{+} e^{-} \rightarrow p \bar{p} J/\psi$ are listed in Fig.~\ref{Fig.2}, where diagrams (a) and (b) correspond to the $P^{(\ast)}_{c}$ and $\bar{P}^{(\ast)}_{c}$ contributions, respectively. With the effective Lagrangians in Eq.~\eqref{eq.2}, one can obtain the amplitudes corresponding to the $e^{+} e^{-} \rightarrow p \bar{p} J/\psi$ process, which are,
\begin{eqnarray}
\mathcal{M}^{\prime 1/2^{-}}_{a}&=&\Big[\bar{u}(p_{4})\Big(g^{1/2^{-}}_{P_{c} \psi N} \gamma_{5} \gamma_{\mu}\Big) \epsilon_{\mu}(p_{5}) \Big] \mathcal{S}^{1/2}(k_{1},m_{P_{c}},\Gamma_{P_{c}}) \nonumber\\ &\times& \Big[\frac{eh}{2m_{N}} \sigma_{\nu \alpha} (ik^{\nu}) v(p_{3})\Big] \frac{-g^{\alpha \beta}}{k^{2}} \Big[\bar{v}(p_{1}) (ie \gamma^{\beta}) u(p_{2}) \Big] \nonumber\\ &\times& F(k,m_{\gamma},\Lambda_{r}) F(k_{1},m_{P_{c}},\Lambda_{r}), \nonumber\\
\mathcal{M}^{\prime 3/2^{-}}_{a}&=&\Big[\bar{u}(p_{4}) \Big( g^{3/2^{-}}_{P_{c} \psi N} \Big) \epsilon_{\mu}(p_{5})\Big] \mathcal{S}^{3/2}_{\mu \tau}(k_{1},m_{P_{c}},\Gamma_{P_{c}}) \nonumber\\ &\times& \Big[\frac{ieh_{1}}{2 m_{N}} \gamma_{\omega} (ik^{\tau}g^{\omega \rho} - ik^{\omega}g^{\tau \rho}) v(p_{3}) \Big] \frac{-g^{\rho \alpha}}{k^{2}} \nonumber\\ &\times& \Big[\bar{v}(p_{1}) (ie \gamma^{\alpha}) u(p_{2}) \Big] F(k,m_{\gamma},\Lambda_{r}) F(k_{1},m_{P_{c}},\Lambda_{r}), \nonumber\\
\mathcal{M}^{\prime 1/2^{-}}_{b}&=&\Big[\bar{u}(p_{4})\Big(g^{1/2^{-}}_{P_{c} \psi N} \gamma_{5} \gamma_{\mu}\Big) \epsilon_{\mu}(p_{5}) \Big] \mathcal{S}^{1/2}(k_{2},m_{P_{c}},\Gamma_{P_{c}}) \nonumber\\ &\times& \Big[\frac{eh}{2m_{N}} \sigma_{\nu \alpha} (ik^{\nu}) v(p_{3})\Big] \frac{-g^{\alpha \beta}}{k^{2}} \Big[\bar{v}(p_{1}) (ie \gamma^{\beta}) u(p_{2}) \Big] \nonumber\\ &\times& F(k,m_{\gamma},\Lambda_{r}) F(k_{2},m_{P_{c}},\Lambda_{r}), \nonumber\\
\mathcal{M}^{\prime 3/2^{-}}_{b}&=&\Big[\bar{u}(p_{4}) \Big( g^{3/2^{-}}_{P_{c} \psi N} \Big) \epsilon_{\mu}(p_{5})\Big] \mathcal{S}^{3/2}_{\mu \tau}(k_{2},m_{P_{c}},\Gamma_{P_{c}}) \nonumber\\ &\times& \Big[\frac{ieh_{1}}{2 m_{N}} \gamma_{\omega} (ik^{\tau}g^{\omega \rho} - ik^{\omega}g^{\tau \rho}) v(p_{3}) \Big] \frac{-g^{\rho \alpha}}{k^{2}} \nonumber\\ &\times& \Big[\bar{v}(p_{1}) (ie \gamma^{\alpha}) u(p_{2}) \Big] F(k,m_{\gamma},\Lambda_{r}) F(k_{2},m_{P_{c}},\Lambda_{r}), \nonumber\\
\hspace{1.5em}
\label{Eq.7}
\end{eqnarray}
where the superscripts $1/2^{-}$ and $3/2^{-}$ have the same meaning as those in Eq.~\eqref{eq.5}. The subscripts $a$ and $b$ correspond to diagrams (a) and (b) of Fig.~\ref{Fig.2}, respectively. In addition, $\mathcal{S}^{1/2}(k_{i},m_{i},\Gamma_{i})$ and $\mathcal{S}^{3/2}_{\mu \nu}(k_{i},m_{i},\Gamma_{i})$ are the propagators of $P_{c}$ states with four-momentum $k_{i}$, mass $m_{i}$ and width $\Gamma_{i}$, respectively, and the specific forms are,
\begin{eqnarray}
&&\mathcal{S}^{1/2}(k_{i},m_{i},\Gamma_{i}) = \frac{\slash\!\!\!k_{i}+m_{i}} {k^2_{i}-m^2_{i} +i m_{i} \Gamma_{i}},\nonumber\\
&&\mathcal{S}^{3/2}_{\mu \nu}(k_{i},m_{i},\Gamma_{i})  = \frac{\slash\!\!\!k_{i}+m_{i}} {k^2_{i}-m^2_{i} +i m_{i} \Gamma_{i}} \nonumber\\&&\qquad \qquad\times  \Big(-g^{\mu \nu} + \frac{\gamma^{\mu} \gamma^{\nu}}{3} + \frac{2 k^{\mu}_{i} k^{\nu}_{i}}{3 m^{2}_{i}} +\frac{\gamma^{\mu} k^{\nu}_{i}-k^{\mu}_{i} \gamma^{\nu}}{3 m_{i}} \Big).\qquad \label{Eq.8}
\end{eqnarray}

\section{NUMERICAL RESULTS AND DISCUSSIONS}
\label{sec:MA}
\subsection{Parameter in the form factor and coupling constants}
The parameter $\Lambda_{r}$ in the form factor is of order 1 GeV, and usually determined by comparing the theoretical estimations with the corresponding experimental data. However, no experimental measurements currently exist for the process of interest. Notably, in analogy to $P_c(4440)$ and $P_c(4457)$, which are interpreted as $D^\ast \Sigma_c$ molecular states, $\Lambda_c(2910)$ and $\Lambda_c(2940)$ have been identified as $D^\ast N$ molecular states with $J^P=1/2^-$ and $3/2^-$, respectively. The production of $\Lambda_{c}(2910)/\Lambda_c(2940)$ has been explored in a variety of  processes, including $p \bar{p} \rightarrow \bar{\Lambda}_{c} \Lambda_{c}$~\cite{Haidenbauer:2009ad}, $p \bar{p} \rightarrow \bar{\Lambda}_{c} \Lambda_{c}(2940)$~\cite{He:2011jp, Dong:2014ksa}, $\pi^{-}p \rightarrow D^{-}D^{0}p$~\cite{Xie:2015zga}, and $\gamma n \rightarrow D^{-} \Lambda_{c}(2940)$~\cite{Wang:2015rda}. In these investigations, the form factors are also employed as the one in Eq.~\eqref{Eq.6} with $\Lambda_r=3.0$ GeV. In general, the cross sections increase monotonically with increasing $\Lambda_r$, so a larger $\Lambda_r$ can be used to estimate the upper limit of the cross sections. In the present work, we vary $\Lambda_r$ from 2 to 3 GeV to check the model dependence of the present estimations. It is worth noting that the authors in Ref.~\cite{Zhang:2025pfz} estimated the cross section for $e^{+} e^{-} \rightarrow p \bar{p} J/\psi$. Their results indicated that the cross section for $e^{+}e^{-} \rightarrow \bar{p} P_{c}$ is $\lesssim \mathcal{O}(0.1~\text{pb})$ around $\sqrt{s}=6$ GeV. Therefore, we can compare our estimations with this upper limit to further test the validity of the parameter range.

In addition to the parameter $\Lambda_{r}$, the values of coupling constants should be determined before estimating the cross sections. For the coupling constants $g^{1/2^{-}}_{P_{c} \psi N}$ and $g^{3/2^{-}}_{P_{c} \psi N}$, using the effective Lagrangians in Eq.~\eqref{eq.2}, one can obtain the corresponding amplitudes $\mathcal{M}_{P_{c}\rightarrow J/\psi p}$. Then, the decay width of the $P_{c}\rightarrow J/\psi p$ process can be written as,
\begin{eqnarray}
\Gamma_{P_{c} \rightarrow J/\psi p} = \frac{1}{(2J+1)8\pi} \frac{|\vec{k}_f|}{M^{2}} \overline{|\mathcal{M}_{P_{c} \rightarrow J/\psi p}|^2}, \label{Eq.9}
\end{eqnarray}
where $M$ and $J$ refer to the mass and angular momentum of the initial $P_{c}$ states, respectively. $\vec{k}_f$ is the three-momentum of the final states in the initial rest frame. 

\begin{table}
\caption{A summary of theoretical estimations  for the branching ratios of $\mathcal{B}(P_{c}\rightarrow J/\psi p)$.}
\label{Tab.a}
\renewcommand{\arraystretch}{2}
\setlength{\tabcolsep}{6pt}
\centering
\begin{tabular}{ccc}
\hline\hline
Framework & $\mathcal{B}(P_{c}\rightarrow J/\psi p)$ & Reference\\
\hline
\multirow{3}{*}{Pentaquark} & $0.10\%<\mathcal{B}(P_{c}(4312)\rightarrow J/\psi p)<2.0\%$ & \multirow{3}{*}{\cite{Cao:2019kst}} \\
& $0.36\%<\mathcal{B}(P_{c}(4440)\rightarrow J/\psi p)<2.0\%$ & \\
& $0.10\%<\mathcal{B}(P_{c}(4457)\rightarrow J/\psi p)<2.0\%$ & \\ \cline{1-3}
\multirow{13}{*}{Molecular} & $30\%<\mathcal{B}(P_{c}(4312)\rightarrow J/\psi p)<78\%$ & \multirow{3}{*}{\cite{Xiao:2019mvs}} \\
& $24\%<\mathcal{B}(P_{c}(4440)\rightarrow J/\psi p)<76\%$ & \\
& $31\%<\mathcal{B}(P_{c}(4457)\rightarrow J/\psi p)<62\%$ & \\ \cline{2-3}
  & $26\%<\mathcal{B}(P_{c}(4312)\rightarrow J/\psi p)<84\%$ & \multirow{3}{*}{\cite{Wu:2019rog}} \\
& $20\%<\mathcal{B}(P_{c}(4440)\rightarrow J/\psi p)<67\%$ & \\
& $23\%<\mathcal{B}(P_{c}(4457)\rightarrow J/\psi p)<79\%$ & \\ \cline{2-3}
  & $0.10\%<\mathcal{B}(P_{c}(4312)\rightarrow J/\psi p)<0.9\%$ & \multirow{3}{*}{\cite{Lin:2019qiv}} \\
& $1.5\%<\mathcal{B}(P_{c}(4440)\rightarrow J/\psi p)<6.0\%$ & \\
& $2.1\%<\mathcal{B}(P_{c}(4457)\rightarrow J/\psi p)<6.1\%$ & \\ \cline{2-3}
  & $2.4\%<\mathcal{B}(P_{c}(4312)\rightarrow J/\psi p)<4.0\%$ & \multirow{3}{*}{\cite{Wang:2019spc}} \\
& $11\%<\mathcal{B}(P_{c}(4440)\rightarrow J/\psi p)<16\%$ & \\
& $1.6\%<\mathcal{B}(P_{c}(4457)\rightarrow J/\psi p)<2.8\%$ & \\ \cline{2-3}
 & $11\%<\mathcal{B}(P_{c}(4312)\rightarrow J/\psi p)<26\%$ & \multirow{1}{*}{\cite{Yang:2024nss}} \\ \cline{2-2}
\hline\hline
\end{tabular} 
\end{table}

The absolute branching fractions for $P_c\to J/\psi p$ have not been experimentally measured yet. On the theoretical side, we summarize the theoretical results for $\mathcal{B}(P_{c}\rightarrow J/\psi p)$ in Table~\ref{Tab.a}. Specifically, the authors in Ref.~\cite{Xiao:2019mvs, Wu:2019rog} studied the $P_{c} \rightarrow J/\psi p$ decays in the molecular frame. Their estimations suggested that the branching fractions of the $J/\psi p$ channel for $P_{c}(4312)/P_{c}(4440)/P_{c}(4457)$ are several tens of percent. Following the same molecular frame, the authors in Ref.~\cite{Wang:2019spc} calculated the decay widths of $P_{c} \rightarrow J/\psi p$ using a quark interchange model. Their results indicated that the branching fractions of the $ J/\psi p$ channel for $P_{c}(4312)/P_{c}(4440)/P_{c}(4457)$ range from a few percent to over ten percent. However, in the pentaquark frame, the authors in Ref.~\cite{Cao:2019kst} estimated that the upper limits of the $J/\psi p$ branching fractions for the three $P_{c}$ states are $2\%$, implying the theoretical predictions of the branching fractions are strongly model dependent. Moreover, in theoretical works on the production of $P_{c}$ states. the authors in Ref.~\cite{Wang:2019krd} estimated the cross sections for the $\gamma p \rightarrow J/\psi p$ process in the molecular frame, where the branching fractions for $P_{c} \rightarrow J/\psi p$ were assumed to be $3\%$ and $10\%$, respectively. In the same molecular frame, our previous work~\cite{Liu:2021ojf} investigated the productions of $P_{c}$ states in association with $Z_{c}(3900)/Z_{cs}(3985)$ via $\pi p$ and $Kp$ scattering, where the branching fractions for $P_{c} \rightarrow J/\psi p$ were assumed to be $10\%$. In the following estimations, we adopt $\mathcal{B}(P_{c}\rightarrow J/\psi p)=10\%$ to estimate the cross sections of considered processes.

The cross sections scale linearly with these branching fractions, allowing us to extract cross sections for other values of the branching fractions through their relative ratios. Using the total widths of $P_{c}(4312)$, $P_{c}(4440)$, and $P_{c}(4457)$ given in Eq.~\eqref{eq.1}, the coupling constants $g^{1/2^{-}}_{P_{c} \psi N}$ and $g^{3/2^{-}}_{P_{c} \psi N}$ can be determined, and subsequently obtain the electromagnetic coupling constants $eh$ and $eh_{1}$. The resulting values are displayed in Table~\ref{Tab.1}. It should be noted that the $\gamma N P_{c}$ and $P_{c} \psi N$ vertices are involved in amplitude of $e^+e^- \to \bar{p} p J/\psi$, leading to an overall quadratic dependence on the $g_{P_c \psi p}^2$ for the amplitudes themselves. Consequently, the interference between $P_{c1}$ and $P_{c2}$ is proportional to $g_{P_{c1} \psi p}^2 g_{P_{c2} \psi p}^2$. Therefore, the signs of the coupling constants listed in Table~\ref{Tab.1} have no impact on the cross sections for the considered processes.

For the decay constant $f_{\psi}$, using the effective Lagrangian in Eq.~\eqref{eq.4}, the decay width of the $J/ \psi \rightarrow e^{+} e^{-}$ process can be estimated as,
\begin{eqnarray}
\Gamma_{J/ \psi \rightarrow e^{+} e^{-}} =  \left(\frac{e}{f_{\psi}} \right)^{2} \frac{8 \alpha |{\vec{p_{e}}}| ^{3}}{3 m^{2}_{\psi}}, \label{Eq.10}
\end{eqnarray}
where $\alpha=1/137$ is the fine-structure constant. $\vec{p_{e}}$ refers to the three-momentum of the electron in the initial rest frame. With the partial decay width $\Gamma_{J/ \psi \rightarrow e^{+} e^{-}}=5.428$ keV~\cite{ParticleDataGroup:2024cfk}, one can obtain ${e}/{f_{\psi}}\backsimeq0.027$.
\begin{table}
\caption{The coupling constants $g_{P_{c} J/ \psi p}$ and electromagnetic coupling constants $eh$ with the $J/ \psi p$ channel accounting for $10\%$ of the total width of each $P_{c}$ state.}
\label{Tab.1}
\renewcommand{\arraystretch}{2}
\setlength{\tabcolsep}{14pt}
\centering
\begin{tabular}{cccc}
\hline\hline
States&$J^{P}$ & $g_{P_{c} J/ \psi p}$&$eh/eh_{1}$\\
\hline
$P_{c}(4312)$&$1/2^{-}$&$0.11$&$0.0026$\\
$P_{c}(4440)$&$1/2^{-}$&$0.14$&$0.0032$\\
$P_{c}(4457)$&$3/2^{-}$&$0.08$&$0.0018$\\
\hline \hline
\end{tabular} 
\end{table}

\begin{figure*}[htb]
     \includegraphics[width=175mm]{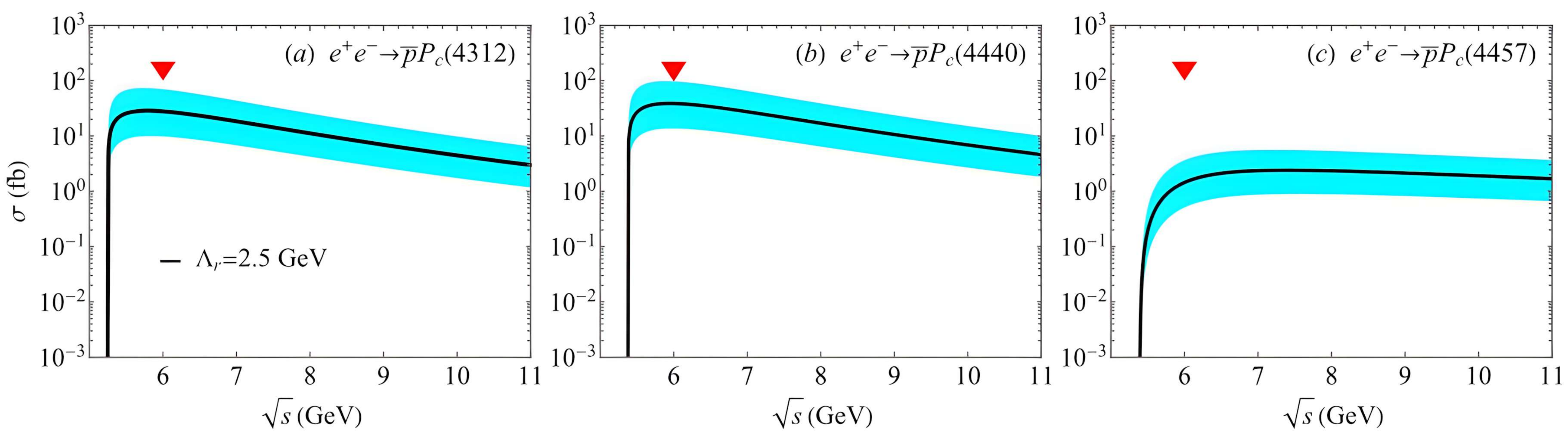}
	\caption{(Color online.) The cross sections for the $e^{+} e^{-} \rightarrow \bar{p} P_{c}(4312)$/$P_{c}(4440)$/$P_{c}(4457)$ processes depending on the center-of-mass energy $\sqrt{s}$. Diagrams (a), (b), and (c) correspond to the contributions from $\bar{p} P_{c}(4312)$, $\bar{p} P_{c}(4440)$, and $\bar{p} P_{c}(4457)$ final states of the relevant processes, respectively. The black solid curves are obtained with $\Lambda_{r}=2.5$ GeV, while the cyan band are the uncertainties resulted from the varying of $\Lambda_{r}$ from 2.0 to 3.0 GeV. The red inverted triangles represent the upper limits of cross section $\sigma \leq 100$ fb at $\sqrt{s}=6$ GeV from Ref.~\cite{Zhang:2025pfz}.
\label{Fig.3}}
\end{figure*}

\begin{figure*}[htb]
     \includegraphics[width=175mm]{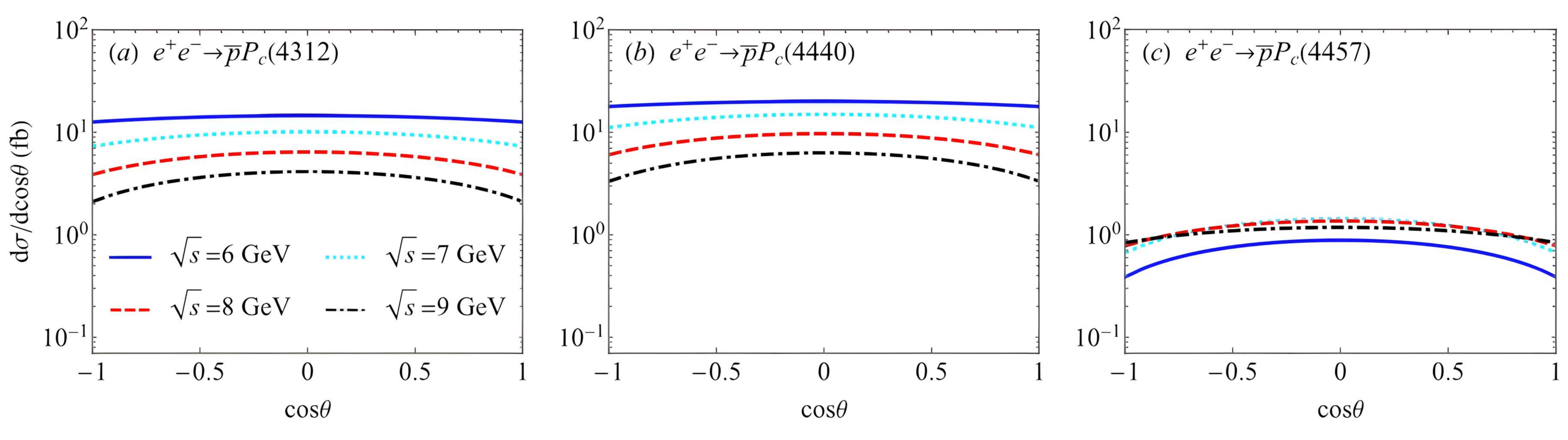}
	\caption{(Color online.) The differential cross sections for the $e^{+} e^{-} \rightarrow \bar{p} P_{c}(4312)$/$P_{c}(4440)$/$P_{c}(4457)$ processes depending on cos$\theta$. Diagrams (a), (b), and (c) correspond to the $\bar{p} P_{c}(4312)$, $\bar{p} P_{c}(4440)$, and $\bar{p} P_{c}(4457)$ final states of the relevant processes, respectively. The parameter $\Lambda_{r}$ is taken to be 2.5 GeV.
\label{Fig.4}}
\end{figure*}

\begin{figure*}[htb]
     \includegraphics[width=170mm]{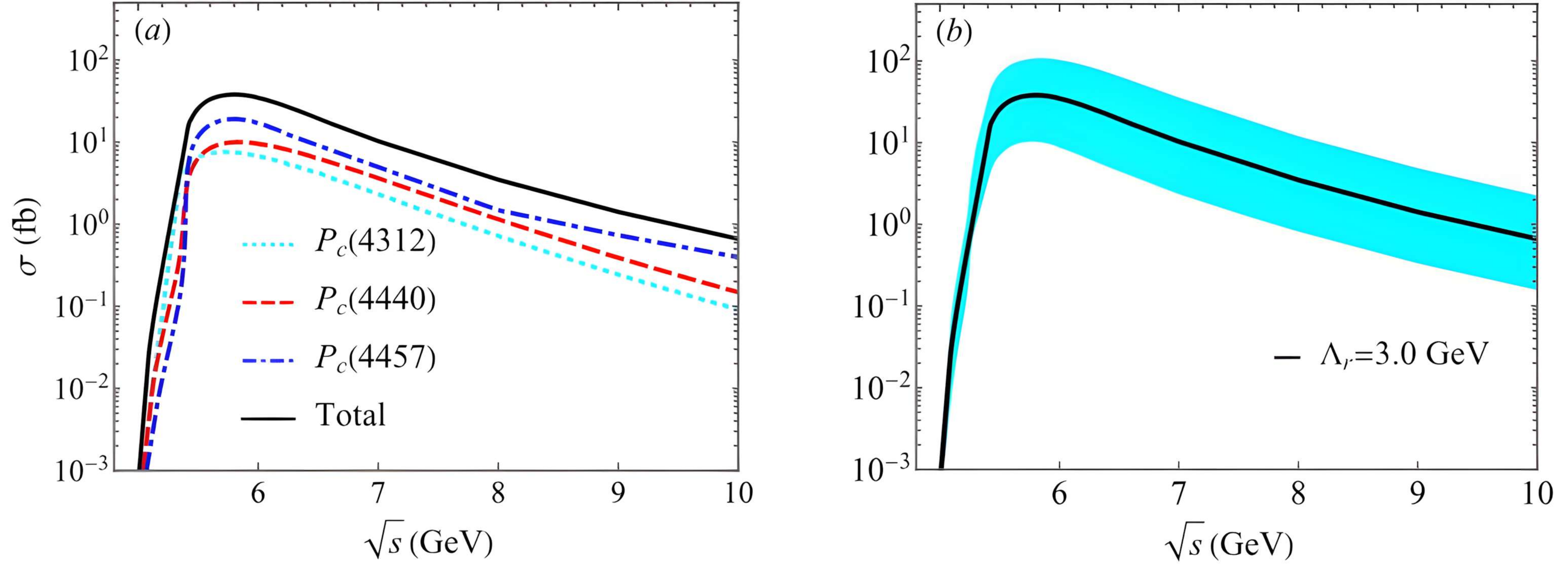}
     \caption{(Color online.) The cross sections for the $e^{+} e^{-} \rightarrow p \bar{p} J/\psi$ process depending on the center-of-mass energy $\sqrt{s}$. Diagram (a) corresponds to the individual contributions from $P_{c}(4312)$, $P_{c}(4440)$, $P_{c}(4457)$, and their corresponding antiparticles, respectively. Diagram (b) corresponds to the total cross sections with the uncertainty resulted from the parameter $\Lambda_{r}$ where the black solid curve is obtained with $\Lambda_{r}=2.5$ GeV, and the cyan band represents the uncertainties resulted from the varying of $\Lambda_{r}$ from 2.0 to 3.0 GeV. The points with error bars correspond to the experimental data of the cross sections for $e^{+} e^{-} \rightarrow p \bar{p} J/\psi$ from Ref.~\cite{Belle-II:2026heb}.
\label{Fig.5}}
\end{figure*}

\subsection{Cross sections for $e^{+} e^{-} \rightarrow \bar{p} P_{c}(4312)$/$P_{c}(4440)$/$P_{c}(4457)$}
With the coupling constants determined in the above subsection, we then estimate the cross sections for the $e^{+} e^{-} \rightarrow \bar{p} P_{c}(4312)$/$P_{c}(4440)$/$P_{c}(4457)$ processes using the amplitudes in Eq.~\eqref{eq.5}. The differential cross sections depending on cos$\theta$ for these two-bdoy processes can be expressed as,
\begin{eqnarray}
\frac{d{\sigma}} {d \cos\theta}
=\frac{1} {32 \pi s} \frac{|\vec{p}_f|} {|\vec{p}_i|}  \left(\frac{1} {4} \left|\overline{\mathcal{M}} \right|^2\right), \label{Eq.11}
\end{eqnarray}
where $s$ refers to the square of the center-of-mass energy. $\theta$ is the scattering angle between the outgoing antiproton and the incoming electron beam direction in the center-of-mass system (CMS). $\vec{p_f}$ and $\vec{p_i}$ refer to three-momenta of the final $P_{c}$ state and the initial electron beam in CMS, respectively.

With the above preparations, the cross sections for $e^{+} e^{-} \rightarrow \bar{p} P_{c}(4312)$/$P_{c}(4440)$/$P_{c}(4457)$ depending on the center-of-mass energy $\sqrt{s}$ are presented in Fig.~\ref{Fig.3}, where diagrams (a), (b), and (c) correspond to the cross sections for $\bar{p} P_{c}(4312)$, $\bar{p} P_{c}(4440)$, and $\bar{p} P_{c}(4457)$ final states, respectively. The red inverted triangles represent the upper limits for the estimated cross sections of $e^{+} e^{-} \rightarrow \bar{p} P_{c}$ from Ref.~\cite{Zhang:2025pfz}. The black solid curves are obtained with $\Lambda_{r}=2.5$ GeV, while the cyan band represents the uncertainties resulted from the variation of $\Lambda_{r}$ from 2.0 to 3.0 GeV. From these three diagrams, one can find that each cross section curve increases rapidly near the threshold. In addition, our results show that the cross sections for $e^+ e^- \to \bar{p} P_{c}(4457)$ process are overall smaller than those for $e^+ e^- \to\bar{p} P_{c}(4312)$ and $\bar{p} P_{c}(4440)$. In particular, for the $e^{+} e^{-} \rightarrow \bar{p} P_{c}(4312)$, $e^{+} e^{-} \rightarrow \bar{p} P_{c}(4440)$, and $e^{+} e^{-} \rightarrow \bar{p} P_{c}(4457)$ processes, the cross sections are estimated to be ($27.8^{+42.2}_{-17.9}$) fb, ($38.7^{+58.6}_{-25.1}$) fb, and ($1.44^{+2.18}_{-0.93}$) fb at $\sqrt{s}=6$ GeV, respectively, which are under the theoretical upper limit estimated in Ref.~\cite{Zhang:2025pfz}. Thus, in the following calculations, we will adopt $\Lambda_{r}=2.5$ GeV to discuss other involved processes.

Besides the cross sections, we also estimate the differential cross sections for $e^{+} e^{-} \rightarrow \bar{p} P_{c}(4312)$/$P_{c}(4440)$/$P_{c}(4457)$ depending on cos$\theta$ with $\Lambda_{r}=2.5$ GeV, which are displayed in Fig.~\ref{Fig.4}. Similarly, diagrams (a), (b), and (c) correspond to the differential cross sections for $e^{+} e^{-}\to \bar{p} P_{c}(4312)$, $e^{+} e^{-}\to \bar{p} P_{c}(4440)$, and $e^{+} e^{-}\to \bar{p} P_{c}(4457)$, respectively. The blue solid, cyan dotted, red dashed, and black dash-dotted curves represent the differential cross sections at $\sqrt{s}=$6, 7, 8, 9 GeV, respectively. Our estimations indicated that these differential cross sections are weakly dependent on the $\cos\theta$. 

\begin{figure*}[htb]
     \includegraphics[width=170mm]{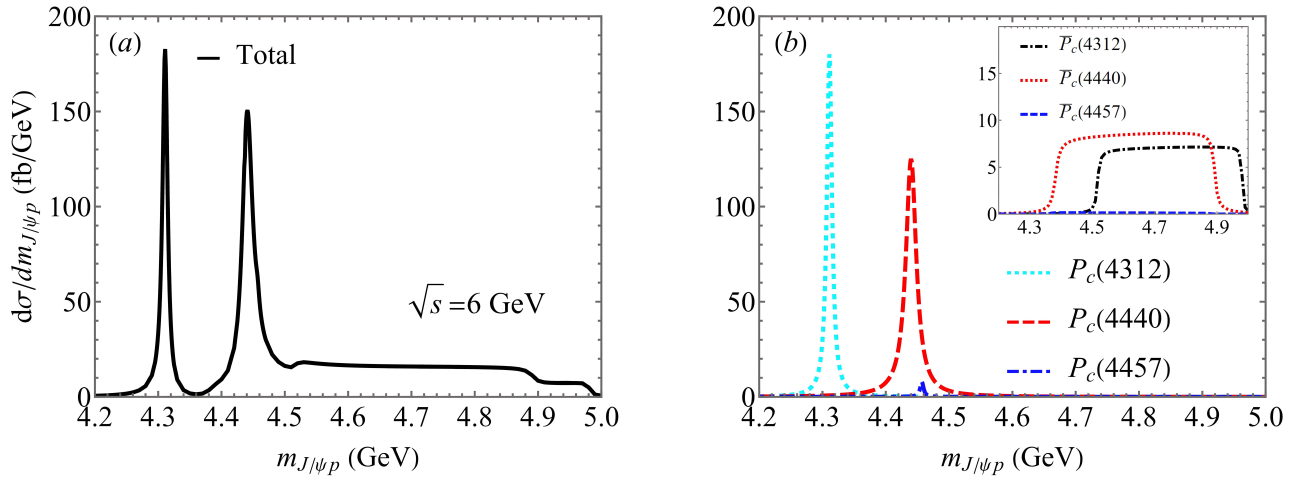}
	\caption{(Color online.) The $J/\psi p$ invariant mass distributions of the $e^{+} e^{-} \rightarrow p \bar{p} J/\psi$ process at $\sqrt{s}=6$ $\mathrm{GeV}$. Diagram (a) corresponds to the total $J/\psi p$ invariant mass spectrum, which represents the summation of the individual contributions from $P_c$ and $\bar{P}_c$ as well as their interferences. Diagram (b) corresponds to the individual contributions from $P_{c}(4312)$, $P_{c}(4440)$, and $P_{c}(4457)$, respectively. In addition, the inner figure corresponds to the individual contributions from $\bar{P}_{c}(4312)$, $\bar{P}_{c}(4440)$, and $\bar{P}_{c}(4457)$, respectively.} 
     \label{Fig.6}
\end{figure*}

\begin{figure*}[htb]
     \includegraphics[width=170mm]{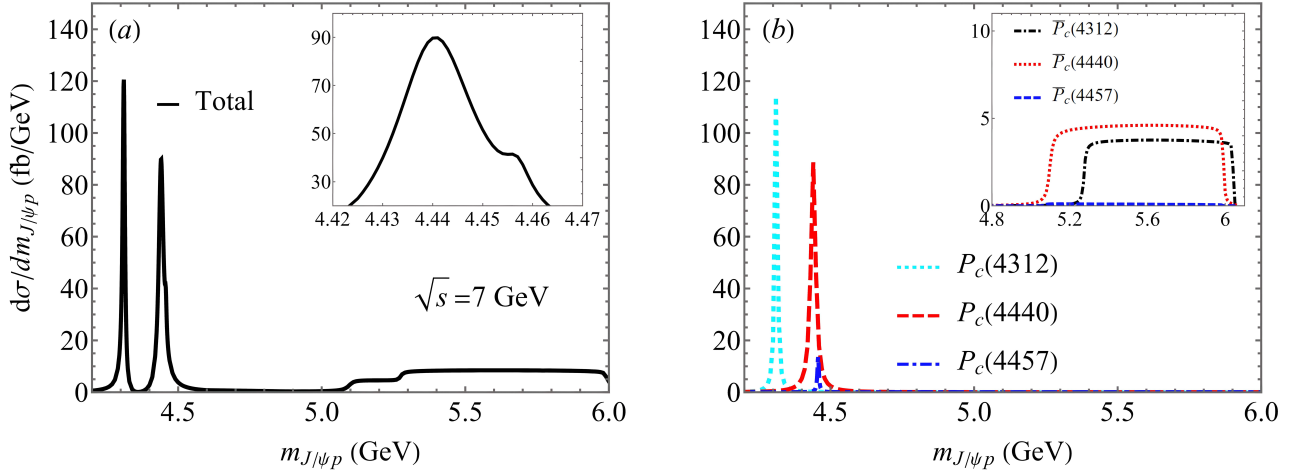}
	\caption{(Color online.) The same as Fig.~\ref{Fig.6} but at $\sqrt{s}=7$ $\mathrm{GeV}$.} 
     \label{Fig.7}
\end{figure*}

\subsection{Cross Sections for $e^{+} e^{-} \rightarrow p \bar{p} J/\psi$}
Since the $P_{c}(4312)$, $P_{c}(4440)$, and $P_{c}(4457)$ states can decay into $J/\psi p$ final states, we further estimate the cross sections for the $e^{+} e^{-} \rightarrow p \bar{p} J/\psi$ process, where the contributions from both $P_c$ and $\bar{P}_c$ states have been included. With the amplitudes in Eq.~\eqref{Eq.7}, the total amplitude of $e^{+} e^{-} \rightarrow p \bar{p} J/\psi$ can be written as,
\begin{eqnarray}
\mathcal{M}^{\prime}_{\mathrm{Tot}}=\mathcal{M}^{\prime 1/2^{-}}_{a}+\mathcal{M}^{\prime 3/2^{-}}_{a}+\mathcal{M}^{\prime 1/2^{-}}_{b}+\mathcal{M}^{\prime 3/2^{-}}_{b}. \label{Eq.12}
\end{eqnarray}

With the above total amplitude, the differential cross sections of the $e^{+} e^{-} \rightarrow p \bar{p} J/\psi$ process can be expressed as,
\begin{eqnarray}
d{\sigma}=\frac{|\vec{p}_M||\vec{p}_{f}|} {128(\pi)^4 \sqrt{s}} \frac{1} {	\Phi} \overline{\left| {\mathcal{M^{\prime}}_{\mathrm{Tot}}}\right|^2 }d\cos\theta_{P_c} d\cos\theta_{J/\psi} d\phi_{J/\psi} dm_{J/\psi p}, \nonumber\\
\label{Eq.13}
\end{eqnarray}
where the flux factor $\Phi=4|{\vec{p_1}}|\sqrt{s}$, $\vec{p_1}$ and $\vec{p}_M$ stand for the three-momentum of the initial electron and the $P_{c}$ state in the $e^{+}e^{-}$ rest frame, respectively. $\vec{p}_f$ is the three-momentum of the $J/\psi$ meson in the $P_{c}$ rest frame. Moreover, the polar angle $\theta_{P_c}$ is the production angle of the $P_{c}$ system with respect to the electron beam direction in the $e^{+}e^{-}$ rest frame,
while $\theta_{J/\psi}$ is the helicity angle between the momentum of $J/\psi$ in the $P_{c}$ rest frame and the $P_{c}$ momentum in the $e^{+}e^{-}$ rest frame, while $\phi_{J/\psi}$ is the corresponding azimuth angle. 

With the above preparations,the cross sections for $e^{+} e^{-} \rightarrow p \bar{p} J/\psi$ depending on $\sqrt{s}$ are presented in Fig.~\ref{Fig.5}. Among them, Fig.~\ref{Fig.5}-(a) corresponds to the cross sections for $e^{+} e^{-} \rightarrow p \bar{p} J/\psi$ estimated with $\Lambda_{r}=2.5$ GeV, where the cyan dotted, red dashed, and blue dash-dotted curves refer to the individual contributions from $P_{c}(4312)$, $P_{c}(4440)$, $P_{c}(4457)$, and their corresponding antiparticles, respectively, while the black solid curve stands for the total cross sections for $e^{+} e^{-} \rightarrow p \bar{p} J/\psi$. For the three individual contributions, one can find that the contributions from $P_{c}(4312)/\bar{P}_{c}(4312)$ exhibit a more rapid increase near the threshold than the other two $P_{c}$ states. As $\sqrt{s}$ increases, our results show that the cross sections resulted from $P_{c}(4312)/\bar{P}_{c}(4312)$ and $P_{c}(4440)/\bar{P}_{c}(4440)$ states reach their maximum near $\sqrt{s}=6$ GeV. Moreover, the cross section of $P_{c}(4440)/\bar{P}_{c}(4440)$ is about 6.8 fb at $\sqrt{s}=7$ GeV, which is about 30 times that from $P_{c}(4457)/\bar{P}_{c}(4457)$. In Fig.~\ref{Fig.5}-(b), we present the total cross sections for $e^{+} e^{-} \rightarrow p \bar{p} J/\psi$, where the black curve is obtained with $\Lambda_{r}=2.5$ GeV, and the cyan band represents the uncertainties resulted from the variation of $\Lambda_{r}$ from 2.0 to 3.0 GeV. In particular, the total cross sections are estimated to be ($14.7^{+22.2}_{-9.50}$) fb at $\sqrt{s}=6$ GeV, indicating that its value spans nearly an order of magnitude in the considered parameter $\Lambda_{r}$ range. In addition, the Belle II Collaboration recently reported the first measurement of the $e^{+} e^{-} \rightarrow p \bar{p} J/\psi$ process from threshold to 7 GeV via initial-state radiation~\cite{Belle-II:2026heb}. In some energy range, no events were observed, and only the upper limits of the cross sections were obtained. In Fig.~\ref{Fig.5}-(b), we present the Belle data with non-zero central values as the blue points with error bars for comparison. The estimated cross sections in the present work are safely below the experimental data.

In addition to the cross sections, the differential cross section for $e^{+} e^{-} \rightarrow p \bar{p} J/\psi$ depending on the $J/\psi p$ invariant mass are presented in Fig.~\ref{Fig.6}. In Fig.~\ref{Fig.6}-(a), the black solid curve represents the summation of the individual contributions from the $P_{c}$ and $\bar{P}_{c}$ states as well as their interferences. In this diagram, a structure near 4.31 GeV is clearly visible. As for $P_{c}(4440)$ and $P_{c}(4457)$, our estimations show that the signal of $P_{c}(4440)$ is predominant, while that of $P_{c}(4457)$ is much smaller. In Fig.~\ref{Fig.6}-(b), the cyan dotted, red dashed and blue dash-dotted curves correspond to the individual contributions from $P_{c}(4312)$, $P_{c}(4440)$, and $P_{c}(4457)$, respectively. From this diagram, three narrow structures can be observed in the $J/\psi p$ invariant mass spectrum near 4.31, 4.44, and 4.46 GeV, corresponding to $P_{c}(4312)$, $P_{c}(4440)$, and $P_{c}(4457)$ states, respectively. On the whole, the signal of $P_{c}(4440)$ is larger than that of $P_{c}(4312)$ and $P_{c}(4457)$. Moreover, the signals of $P_{c}(4440)$ and $P_{c}(4457)$ are partially overlap. As shown in the inner diagram of Fig.~\ref{Fig.6}-(b), the black dash-dotted, red dotted, and bule dashed curves correspond to the individual contributions from $\bar{P}_{c}(4312)$, $\bar{P}_{c}(4440)$, and $\bar{P}_{c}(4457)$, respectively. In contrast to the $P_{c}$ states, the signals of the $\bar{P}_{c}$ states are rather broad in the $J/\psi p$ invariant mass spectrum, while their contributions overlap in the $m_{J/\psi p}$ range from 4.3 to 5.0 GeV.

Considering that the phase space increases with the increase of center-of-mass energy, we further present the $J/\psi p$ invariant mass distributions of the $e^{+} e^{-} \rightarrow p \bar{p} J/\psi$ process at $\sqrt{s}=7$ $\mathrm{GeV}$ in Fig.~\ref{Fig.7}. In this diagram, the signals near 4.31 GeV and 4.45 GeV in the $J/\psi p$ invariant mass spectrum are visible. As shown in the inner figure of Fig.~\ref{Fig.7}-(a), a clear structure is observed between 4.42 and 4.46 GeV in the $J/\psi p$ invariant mass spectrum, which should correspond to $P_{c}(4440)$ rather than $P_{c}(4457)$. Moreover, a weak signal from $P_{c}(4457)$ is also visible between 4.45 GeV and 4.46 GeV. In addition, the overlap between the $P_{c}$ and $\bar{P}_{c}$ states is negligible, which suggests that the $P_{c}$ states can be experimentally probed at this energy point without significant interference from the $\bar{P}_{c}$ states. In Fig.~\ref{Fig.7}-(b), one can find that the signal strengths of the $P_{c}$ and $\bar{P}_{c}$ states are a bit smaller than those in Fig.~\ref{Fig.6}-(b), and the reflections of $\bar{P}_c$ states in the $J/\psi p$ invariant mass distributions is above 5 GeV, which do not overlap with the contributions from $P_c$ states.

\begin{figure*}[htbp]
     \includegraphics[width=170mm]{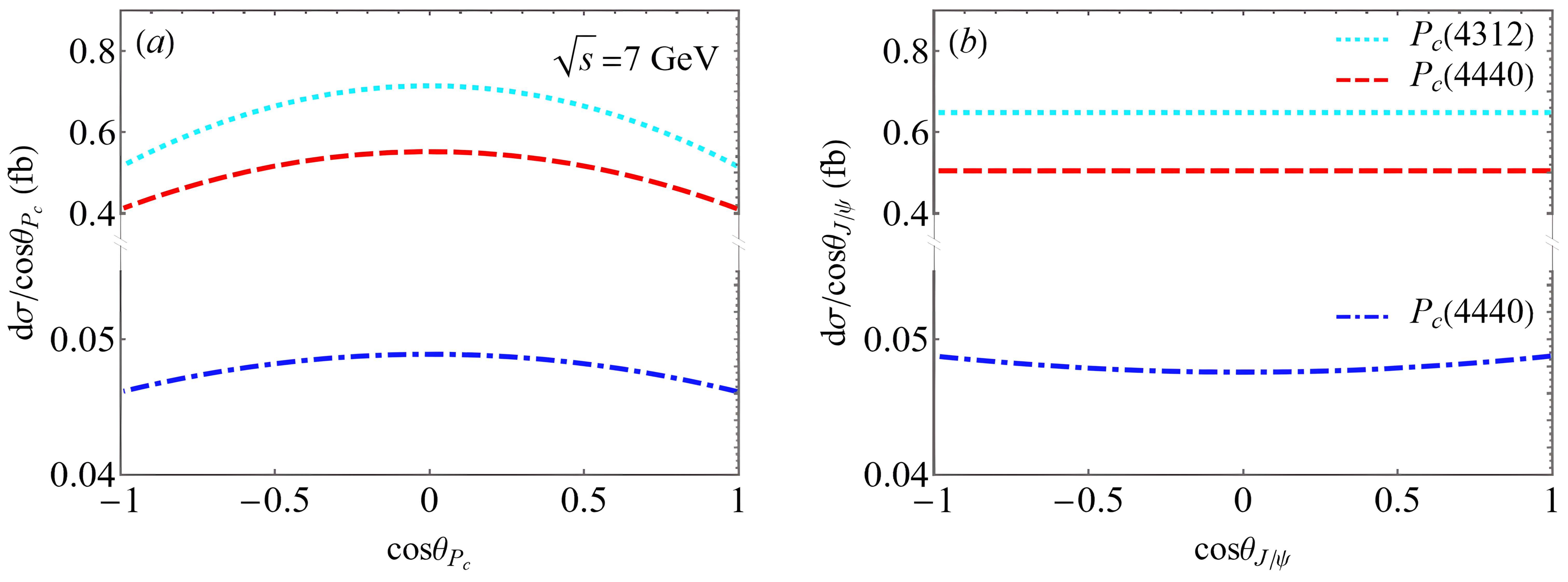}
     \caption{(Color online.) The polar angle $\theta_{P_{c}}$ distribution (Diagram (a)) and helicity angle $\theta_{J/\psi}$ distribution (Diagram (b)) for the $e^{+} e^{-} \rightarrow p \bar{p} J/\psi$ process at $\sqrt{s}=7$ GeV. The cyan dotted, red dashed, and blue dash-dotted curves refer to the individual contributions from $P_{c}(4312)$, $P_{c}(4440)$, and $P_{c}(4457)$, respectively.} 
     \label{Fig.8}
\end{figure*}

In the present work, we assign $J^{P}=1/2^{-}$ to $P_{c}(4312)$ and $P_{c}(4440)$, while $J^{P}=3/2^{-}$ to $P_{c}(4457)$. Considering that the helicity angle distribution is the best quantity to identify the quantum numbers of the $P_{c}$ states, we estimate the polar angle $\theta_{P_{c}}$ distribution  and helicity angle $\theta_{J/\psi}$ distribution for $e^{+} e^{-} \rightarrow p \bar{p} J/\psi$ at $\sqrt{s}=7$ GeV. To separate the contributions from different $P_c$ states, we choose $m_{J/\psi p}  \in [m_{P_c}-\Gamma_{P_c},m_{P_c}+\Gamma_{P_c}]$ for $P_{c}(4312)$ and $P_c(4457)$. As for $P_c(4440)$, we choose $m_{J/\psi p} \in  [4433.6,4446.4]$ MeV to avoid the mixing between $P_c(4440)$ and $P_c(4457)$. In Fig.~\ref{Fig.8}, we present the polar angle $\theta_{P_{c}}$ distribution (diagram (a)) and helicity angle $\theta_{J/\psi}$ distribution (diagram (b)) for $e^{+} e^{-} \rightarrow p \bar{p} J/\psi$ at $\sqrt{s}=7$ GeV, where the cyan dotted, red dashed, and blue dash-dotted curves refer to the individual contributions from $P_{c}(4312)$, $P_{c}(4440)$, and $P_{c}(4457)$, respectively. In Fig.~\ref{Fig.8}-(a), our estimations indicate that the polar angle $\theta_{P_{c}}$ distributions for the three $P_{c}$ states reach their maxima at cos$\theta_{P_{c}}=0$ and drop to their minima at cos$\theta_{P_{c}}=\pm 1$. In Fig.~\ref{Fig.8}-(b), one can find that the helicity angle $\theta_{J/\psi}$ distributions for $P_{c}(4312)$ and $P_{c}(4440)$ are isotropic, appearing as straight lines. By contrast, the distribution for $P_{c}(4457)$ shows a slightly concave shape, with a maximum at cos$\theta_{J/\psi}=\pm 1$ and minima at cos$\theta_{J/\psi}=0$, which could be described by $A +B\cos^2\theta$-like dependence. Thus, the two types of distributions are distinguishable, providing a powerful experimental method for determining the quantum numbers of the $P_{c}$ states at future facilities like the STCF.

Before concluding this work, it is worth noting that the STCF is designed to operate with a center-of-mass energy range of 2 to 7 GeV and a peak luminosity of 5.0 $\times$ $10^{34}$ $\mathrm{cm}^{-2}\mathrm{s}^{-1}$, which translates to a yield of events more than 50 times greater than that of BEPCII~\cite{Peng:2020orp, Achasov:2023gey, Ai:2025xop}. Assuming that the STCF can accumulate data for 200 days per year, and using our estimated cross sections for $e^{+} e^{-} \rightarrow p \bar{p} J/\psi$ at $\sqrt{s}=6$ GeV, we find that the STCF can yield $(1.3^{+1.9}_{-0.8}) \times 10^{4}$ $J/\psi p \bar{p}$ events per year. The above estimation is based on a branching fraction of $10\%$ for $P_c\to J/\psi p$. Furthermore, if we assume a branching fraction of $3\%$ for $P_c\to J/\psi p$, the coupling constants $g_{P_{c} J/ \psi p}$ for $P_{c}(4312)$, $P_{c}(4440)$, and $P_{c}(4457)$ are estimated to be 0.06, 0.08, and 0.04, respectively. In this case, our estimates suggest an event yield on the order of $10^{3}$ at STCF.

\section{SUMMARY}
\label{sec:NR}
In 2015, the LHCb Collaboration observed $P_{c}(4380)$ and $P_{c}(4450)$ in the $J/\psi p$ invariant mass distributions of the $\Lambda^{0}_{b} \rightarrow J/\psi K^{-} p$ decay. Four years later, $P_{c}(4312)$ was identified in the same process, while the previously observed $P_{c}(4450)$ structure was resolved into two separate narrow states, $P_{c}(4440)$ and $P_{c}(4457)$. The observed masses of $P_{c}(4312)$ and $P_{c}(4440)$/$P_{c}(4457)$ lie close to the thresholds of $\Sigma_{c}\bar{D}$ and $\Sigma_{c}\bar{D}^{\ast}$, respectively, strongly favoring the $\Sigma_{c}\bar{D}$ and $\Sigma_{c}\bar{D}^{\ast}$ molecular interpretations. Beyond the mass spectrum and decay properties, the production properties of $P_{c}$ states also offer valuable insight into their inner structure. In the present work, we investigate the $P_{c}(4312)$, $P_{c}(4440)$, and $P_{c}(4457)$ productions via $e^{+} e^{-}$ collisions, a process that can be studied at the STCF.

For the $e^{+} e^{-} \rightarrow \bar{p} P_{c}(4312)$/$P_{c}(4440)$/$P_{c}(4457)$ processes, our estimations show that each cross section curve increases rapidly near the threshold. In comparison, the cross sections for the $e^{+} e^{-} \rightarrow \bar{p} P_{c}(4457)$ process are smaller than those for $e^{+} e^{-} \rightarrow \bar{p} P_{c}(4312)/\bar{p} P_{c}(4440)$ processes. In particular, for the $e^{+} e^{-} \rightarrow \bar{p} P_{c}(4312)$, $e^{+} e^{-} \rightarrow \bar{p} P_{c}(4440)$, and $e^{+} e^{-} \rightarrow \bar{p} P_{c}(4457)$ processes, the cross sections are estimated to be ($27.8^{+42.2}_{-17.9}$) fb, ($38.7^{+58.6}_{-25.1}$) fb, and ($1.44^{+2.18}_{-0.93}$) fb at $\sqrt{s}=6$ GeV, respectively, where the central values are estimated with $\Lambda_{r}=2.5$ GeV, and the uncertainties resulting from the variation of $\Lambda_{r}$ from 2.0 to 3.0 GeV. In addition, for the differential cross sections, our results indicate that the differential cross sections are weakly dependent on the $\cos\theta$.

Since $P_{c}(4312)$, $P_{c}(4440)$, and $P_{c}(4457)$ can decay into $J/\psi p$, we propose to investigate these three $P_{c}$ states in the $e^{+} e^{-} \rightarrow p \bar{p} J/\psi$ process, including the contributions from both $P_c$ states and their antiparticles. For the individual contributions, our estimations indicate that the cross section resulted from the $P_{c}(4440)/\bar{P}_{c}(4440)$ intermediate states is dominant. In particular, we estimate the total cross section for $e^{+} e^{-} \rightarrow p \bar{p} J/\psi$ to be ($14.7^{+22.2}_{-9.50}$) fb at $\sqrt{s}=6$ GeV, which varies by an order of magnitude over the $\Lambda_{r}$ range from 2.0 to 3.0 GeV. In addition to the cross sections, the differential cross sections for $e^{+} e^{-} \rightarrow p \bar{p} J/\psi$ depending on $J/\psi p$ invariant mass are estimated at $\sqrt{s}=6$ GeV. Our results show that a narrow structure can be observed in the $J/\psi p$ invariant mass spectrum near 4.31 GeV, corresponding to the $P_{c}(4312)$ state. Moreover, the dominant contribution to the structure between 4.42 and 4.46 GeV comes from $P_{c}(4440)$, indicating that the expected structure in the $J/\psi p$ invariant mass spectrum should correspond to $P_{c}(4440)$ rather than $P_{c}(4457)$.

\section*{ACKNOWLEDGMENTS}
The authors would like to thanks Dr. Yin Huang for useful discussions. This work is partly supported by the National Natural Science Foundation of China under the Grant Nos. 12175037 and 12335001, as well as supported, in part, by National Key Research and Development Program under the contract No. 2024YFA1610503. Quan-Yun Guo is also supported by the SEU Innovation Capability Enhancement Plan for Doctoral Students (Grant No. CXJH SEU 26160).

\end{document}